\documentclass[10pt,preprint]{aastex}
\shorttitle{New Disks and Jets in Orion}
\shortauthors{Smith et al.}
\begin{document}

%\title{Silhouette Disks with Reflection Nebulae and Outlfows in the
%Orion Nebula and M~43\altaffilmark{1}}
\title{NEW SILHOUETTE DISKS WITH REFLECTION NEBULAE AND OUTFLOWS IN THE
ORION NEBULA AND M43\altaffilmark{1}}

\author{Nathan Smith\altaffilmark{2}, John Bally, Daniel Licht, and
Josh Walawender}
 
\affil{Center for Astrophysics and Space Astronomy, University of
Colorado, 389 UCB, Boulder, CO 80309}

\altaffiltext{1}{Based on observations made with the NASA/ESA {\it
Hubble Space Telescope}, obtained at the Space Telescope Science
Institute, which is operated by the Association of Universities for
Research in Astronomy, Inc., under NASA contract NAS5-26555.}

\altaffiltext{2}{Hubble Fellow; nathans@casa.colorado.edu}

\begin{abstract}

We report the detection of several new circumstellar disks seen in
silhouette against background nebular light in the outskirts of the
Orion nebula and the neighboring H~{\sc ii} region M43.  These were
detected as part of our H$\alpha$ survey of Orion with the Advanced
Camera for Surveys on-board the {\it Hubble Space Telescope}.  Several
of the disks show bipolar reflection nebulae, microjets, or pronounced
temporal variability of their central stars.  The relatively large
fraction of bipolar reflection nebulae or microjets in our sample may
be a selection effect caused by the faint nebular background far from
the Trapezium.  Two disks in our sample are large and particularly
noteworthy: A nearly edge-on disk, d216-0939, is located several
arcminutes northwest of M43 and resembles the famous HH~30 disk/jet
system in Taurus.  It drives the 0.15 pc long bipolar outflow HH 667,
and exhibits a remarkable asymmetric reflection nebula caused by the
tilt of the flared disk.  With a diameter of $\sim$2$\farcs$6 (1200
AU), d216-0939 is as large as the giant edge-on silhouette disk
d114-426 in the core of the Orion Nebula.  The large disk d253-1536 is
located in a binary system embedded within an externally-ionized giant
proplyd in M43.  The disk exhibits distortions which we attribute to
tidal interactions with the companion star.  The bipolar jet HH 668
emerges from the proplyd ionization front in a direction orthogonal to
the disk, and can be traced to the young star embedded within it.  A
bow shock lies 54$\arcsec$ south of this binary system along the
outflow axis.  Proper motions over a 1.4 yr baseline confirm that
these emission knots are indeed moving away from d253-1536, with
speeds as high as $\sim$330 km s$^{-1}$ in the HH~668 microjet, and
slower motion farther from the star.

\end{abstract}


\section{INTRODUCTION}

The {\it Hubble Space Telescope} ({\it HST}) has produced remarkable
images of circumstellar disks surrounding young stars.  By far the
largest population of known circumstellar disks is located in the
Orion nebula at a distance of 460 pc (see Bally et al.\ 2000), where
the disks have typical diameters of 50-1000 AU.  O'Dell, Wen, \& Hu
(1993) found several disks seen in silhouette against the background
nebular light, and dozens of young stellar objects surrounded by
teardrop- or tadpole-shaped ionization fronts.  These are
protoplanetary disks seen purely in silhouette, or externally-ionized
photo-ablating protoplanetary disks (``proplyds''), both rendered
visible by their location in or near an H~{\sc ii} region (O'Dell et
al.\ 1993, 1996; McCullough et al.\ 1995; Bally et al.\ 1998, 2000).
McCaughrean \& O'Dell (1996) found six additional disks seen purely in
silhouette against the bright nebular emission from the Orion nebula.
More detailed {\it HST} imaging studies of the nebular core revealed a
total of 15 silhouette disks (Bally, McCaughrean, \& O'Dell 2000;
McCaughrean et al.\ 1998) and nearly 200 bright proplyds (O'Dell et
al.\ 1993; O'Dell \& Wong 1996; Bally et al. 1998, 2000).  Some
silhouette disks are seen embedded inside bright proplyds (Bally et
al.\ 2000), where in some cases the disks glow in [O~{\sc i}] and
H$_2$ emission (Chen et al.\ 1998; St\"{o}rzer \& Hollenbach 1998).
In addition, several one-sided ``microjets'' are associated with
proplyds (Bally et al.\ 2000; Reipurth \& Bally 2001), indicating that
the embedded stars are still surrounded by active accretion disks.

So far, only the inner portion of the Orion nebula -- a region that
contains about 400 of the nearly 2,000 members of the extended
Trapezium cluster of low mass stars -- has been completely surveyed
with {\it HST}.  However, O'Dell (2001) presented HST images of
several objects outside the nebular core, demonstrating that
additional young stars, disks, and jets remain to be discovered in the
outer regions of the Orion nebula.

With the new Wide Field Camera of the Advanced Camera for Surveys
(ACS/WFC) installed on {\it HST} during the 2002 servicing mission, we
explored the outer Orion Nebula where radiation fields are less
intense than in the nebular core.  The outskirts of the Orion nebula
contain young stars bathed in radiation fields that may be more
typical of the environment in which most stars in the sky are born.
The survey samples a large portion of the Orion Nebula Cluster.  Our
ACS/WFC survey also included the first comprehensive {\it HST} imaging
of the neighboring H~{\sc ii} region M43, several acrminutes north of
the Trapezium, powered by the single B0-type star NU Ori.

The new disks are seen despite the the fainter background nebular
emission far from the Trapezium.  On the other hand, the faint nebular
background probably makes it easier to see extended emission from
bipolar reflection nebulae and microjets emerging from the silhouette
disks --- phenomena which are apparently less common among pure
silhouette disks in the inner Orion nebula (Bally et al.\ 2000).
After a general description of these new silhouette disks (\S 3), we
provide more detailed investigations of two particularly interesting
disks and their associated jets (\S4 and \S 5).  One additional
proplyd with a bipolar jet and silhouette disk, d181-826 and HH~540,
is also seen in our survey in the outer parts of the Orion nebula, but
has been discussed separately (Bally et al.\ 2004) because of its
remarkable morphology.

\section{OBSERVATIONS}

The new ACS/WFC observations reported here were extracted from images
obtained as part of an H$\alpha$ survey of the Orion Nebula and M43
during {\it HST} Cycle 12, using the F658N filter to cover 415 square
arcminutes.  The narrowband F658N filter transmits both H$\alpha$ and
[N~{\sc ii}] $\lambda$6583 in the nebula, although for simplicity we
refer to it as an H$\alpha$ filter.  Further details of the Cycle 12
observations and data reduction procedures are given in an earlier
paper in this series (Bally et al.\ 2004), and are not repeated here.

In addition to these Cycle 12 observations, we obtained ACS/WFC images
at four pointings during Cycle 11, using the F658N filter, the F660N
filter which only transmits [N~{\sc ii}] $\lambda$6583, the F502N
filter sampling [O~{\sc iii}] $\lambda$5007, and the F550M continuum
filter.  These Cycle 11 data, obtained on 2002 August 29 and 30, led
to the discovery of the d216-0939 silhouette disk (see \S 4), which
was not observed during Cycle 12, and for measuring proper motions in
the HH~668 jet associated with d253-1536, which was observed in both
Cycle 11 and 12 (see \S 5). The Cycle 11 observation strategy differed
in minor ways from the Cycle 12 program.  For each pointing in the
F658N filters during Cycle 11, two exposures with a duration of 500
seconds each were obtained with a two-point dither pattern in which
the second exposure was offset from the first by 5$\arcsec$ toward
both the east and the south.  This two-point dither pattern served to
fill the gap between the two 2048$\times$4096 pixel CCD arrays of the
ACS/WFC, and in the fields which were common to both exposures served
to identify cosmic ray hits.  Table~1 lists the central coordinates
and other parameters of the Cycle 11 ACS/WFC observations.

%%%%% SECTION 3
\section{NEW SILHOUETTE DISKS}

Figure 1 shows ten H$\alpha$ images of silhouette disks extracted from
our ACS/WFC survey of Orion; each frame is 7\arcsec$\times$7\arcsec.
The name of each disk follows the coordinate-based naming convention
of O'Dell \& Wen (1994), with the prefix ``d'' for disk.  A few
objects show evidence for either bipolar ``microjets'' (see Reipurth
\& Bally 2001; Bally et al.\ 2000) or reflection nebulae.  Since we
obtained only narrowband H$\alpha$ images for most objects, spectra
are needed to distinguish reflection nebulae from jets. Coordinates
and basic properties of each object are summarized in Table 2, and a
brief description of each disk is given below.

{\bf d053-717} (Figure 1$a$).  A small silhouette disk with a bright
central star is seen roughly 5$\arcmin$ southwest of the Trapezium
outside the bright Huygens region.  The outer edge of the silhouette
disk has a diameter of $\sim$0$\farcs$9 (410 AU).  The disk has a
length-to-width ratio of roughly 5 suggesting a high inclination, but
it cannot be purely edge-on since a bright central star is visible.
Thus, the inclination angle is probably of order
75$\arcdeg\pm$10$\arcdeg$.  No sign of a bright proplyd ionization
front is seen.

{\bf d110-3035} (Figure 1$b$). This unusual object appears to be a
nearly edge-on disk with a bright bipolar reflection nebula.  An
edge-on dark disk much like some other disks seen by {\it HST}
(Padgett et al.\ 1999; Stapelfeldt et al.\ 2003) bisects the small
nebula along P.A.$\simeq$0$\arcdeg$.  The extended portions of the
disk are difficult to see against the faint background nebula, with a
diameter of roughly 0$\farcs$9 (410 AU) or more.  Two emission or
reflection knots are distributed nearly symmetrically on either side
of the disk along an east/west axis perpendicular to the disk.

{\bf d124-132} (Figure 1$c$).  A bright proplyd with a small disk
inside is found about 1$\farcm$5 north of the Trapezium.  The proplyd
ionization front has dimensions of 1$\farcs$2$\times$2$\arcsec$, with
the brightest part of the front facing the Trapezium.  The disk has a
diameter of $\sim$0$\farcs$5 (230 AU) with a likely inclination angle
of $\sim$75$\arcdeg$, and shows faint emission from a bipolar
reflection nebula or microjet emerging perpendicular to the major axis
of the disk.  This polar emission is brighter on the western side of
the disk, suggesting that the western part of the polar axis is tilted
toward us out of the plane of the sky.  A faint protrusion in the
proplyd on the western rim along the disk axis may mark the location
of a shock powered by a jet.

{\bf d132-042} (Figure 1$d$).  d132-042 has a similar size and
inclination as the previous disk, d124-132, it is also embedded in a
bright proplyd, and is located near to it on the sky.  It has a
bipolar microjet that is very faint on the northern side of the disk,
but is bright and extends at least 0$\farcs$4 to the south and
intrudes into the disk on the southern side (thus, this is probably
the blueshifted part of the jet).  There is another bright knot along
the presumed jet axis 0$\farcs$7 south of the source, but this also
appears to be the location of the surrounding proplyd ionization
front.

{\bf d132-1832} (Figure 1$e$).  Bally et al.\ (2000) discovered this
large silhouette disk, located north of the Trapezium near the edge of
the bright inner Orion Nebula.  It is discussed here because our
ACS/WFC H$\alpha$ image provides new information in two respects.
First, the new image has higher resolution and sensitivity than
earlier Wide Field Planetary Camera 2 (WFPC2) images, and shows that
faint translucent material in the outer parts of the disk is
apparently getting swept-back (to the north) by radiation pressure or
a large-scale wind from the Trapezium, located toward the south.
Second, the previous WFPC2 image obtained in Cycle 6 (1998 Apr 4)
shows a very bright central star (see the inset of Figure 1$d$), while
the star has completely faded from visibility in our new ACS image.
The F656N filter of WFPC2 also samples H$\alpha$ but is narrower than
the F658N filter of ACS (i.e. the ACS filter contains more continuum
light).  Therefore, the central star is significantly variable, and it
is unclear whether the change is caused by continuum or line emission.
If the excess emission in the earlier WFPC2 image in 1998 is due to an
H$\alpha$ flare, then this emission line's intrinsic luminosity in the
central source decreased by at least $\sim$4$\times$10$^{27}$ ergs
s$^{-1}$ over a period of only 5 years.

{\bf d141-1952} (Figure 1$f$).  The dark region between M42 and M43
harbors a star surrounded by a small disk seen in silhouette against
the background nebular light.  The disk has a major-axis diameter of
roughly 0.7\arcsec\ and a minor-axis diameter of about 0.6\arcsec\
(320$\times$275 AU), implying that its polar axis is tilted about
55-60\arcdeg\ out of the plane of the sky, with the polar axis
projected on the sky along a southeast/northwest direction. This disk
is seen purely in silhouette, with no associated proplyd ionization
front.

{\bf d216-0939} (Figure 1$g$).  This giant silhouette disk and
reflection nebula is located roughly 7$\arcmin$ north-northwest of M43
in outer reaches of the Orion Nebula with very faint background
nebular emission.  This remarkable object is discussed in more detail
in \S 4.

{\bf d253-1536} (Figure 1$h$).  This large silhouette disk is unique
in that it is apparently part of a binary system embedded within a
giant proplyd in M43. This unusual disk and its associated proplyd and
bipolar jet are discussed in detail in \S 5.

{\bf d280-1720} (Figure 1$i$).  The southwestern part of M43 contains
a star surrounded by a small disk seen in silhouette against the
background nebular light.  The disk has a major-axis diameter of
roughly 0.8\arcsec\ and a minor-axis diameter of about 0.6\arcsec\
(370$\times$275 AU), implying that its polar axis is tilted about
45-50\arcdeg\ out of the plane of the sky.  The disk axis faces
roughly toward P.A.$\simeq$280\arcdeg .  However, the orientation and
inclination can only be measured to about 10\% accuracy due to the
presence of the central star whose point spread function (PSF) makes
the determination of the disk's minor diameter difficult.  This disk
is seen purely in silhouette, with no associated proplyd ionization
front.

{\bf d347-1535} (Figure 1$j$).  This small disk is located about
1$\arcmin$ north/northeast of NU Ori in M43.  The disk is nearly
edge-on, with an inclination of perhaps $\sim$80$\arcdeg$.  The disk
diameter is roughly 0$\farcs$7 (320 AU) with the minor axis oriented
along P.A.$\simeq$45$\arcdeg$=225$\arcdeg$.  It has no identifiable
central star, but does show clear evidence for a remarkable bipolar
microjet or reflection nebula emerging perpendicular to the disk.
Although the northeast part of the jet is brighter, the southwest
portion intrudes farther into the disk perimeter, suggesting that this
part of the jet or reflection nebula is tilted toward us.  There is no
proplyd ionization front that can be associated with this silhouette
disk.

\section{THE REFLECTION NEBULA OF d216-0939 AND THE HH 667 JET}

This nearly edge-on disk is seen in silhouette against the very faint
diffuse nebular light north of the Orion nebula and northwest of M43.
A star and compact reflection nebula occupy the middle of the disk,
while a more diffuse reflection nebula is illuminated primarily on its
eastern face.  A bipolar jet emerges orthogonal to the disk, and faint
emission nebulae farther from the star constitute part of a larger
bipolar flow, HH 667.  This disk/jet system is discussed below.

\subsection{Morphology of the Disk and Reflection Nebula}

Figure 2 shows a more detailed view of d216-0939 than is given in
Figure 1$g$.  We deconvolved the original Cycle 11 images in each
filter using 5 iterations of the {\sc lucy} task in
IRAF,\footnote{IRAF is distributed by the National Optical Astronomy
Observatories, which are operated by the Association of Universities
for Research in Astronomy, Inc., under cooperative agreement with the
National Science Foundation.} with observations of a bright star about
5$\arcsec$ away adopted for the nominal PSF.  With only 5 iterations,
these deconvolved images did not yet have the signature artifacts from
over-processing (rings were seen around stars after about 15
iterations), and the resulting effective FWHM spatial resolution we
achieved was $\sim$0$\farcs$06.  In the 3-color image in Figure 2$a$,
blue or white features are dominated by scattered starlight, while
features that appear red/orange are dominated by H$\alpha$ and [N~{\sc
ii}] emission.

%% silhouette - optical depth - disk size ASR 41
Even though this object is seen against the extremely faint outer
parts of the Orion nebula, the {\it HST} image shows an unambiguous
silhouette disk with a polar axis oriented at roughly
83$\arcdeg$=263$\arcdeg$.  The disk is shown best in the contour plot
of H$\alpha$ in Figure 2$b$, where the blue-tinted contours trace
extinction of background light.  It is clear from this deconvolved
image that the disk cuts all the way across the reflection nebula,
indicating that the apparent bridge between the east and west sides in
Figure 1$g$ was caused by the PSF of {\it HST}.  The deepest
extinction in the disk reaches down to $\sim$20--30\% of the
background nebula, corresponding to an optical depth of $\sim$1.4 at
6563 \AA\ and a grain column density of order 2$\times$10$^{-4}$ g
cm$^{-2}$, with standard assumptions about the grain
properties\footnotemark\footnotetext{Here we have assumed $Q_{\rm
abs}$=0.08 for $a$=0.1 $\micron$ grains at a wavelength of 6563 \AA,
and other typical grain properties to calculate the column mass (see
Smith et al.\ 2004).}.  Integrated over the surface of the disk, this
would require a dust mass of order 1 $M_{\earth}$, but of course this
is only a lower limit since there could also be H$\alpha$ emission
along our line of sight to the disk, or additional mass hidden in
denser optically-thick and unresolved parts of the disk.

The diameter of the disk is at least 2$\farcs$6 (1200 AU), but could
be somewhat larger if there is a more diffuse component at larger
radii that is difficult to detect against the very faint nebular
background.  The clear extinction of background light by the outer
edge of the silhouette disk gives a reliable estimate of the disk
diameter, and means that it is not simply a shadow cast by a smaller
disk, as in the bipolar reflection nebula around the young star ASR~41
in NGC~1333 (Hodapp et al.\ 2004).  The gap between the east and west
sides of the reflection nebula at 1$\arcsec$--1$\farcs$8 south of the
central star may be this type of shadow, or it may indicate a larger
diffuse disk not seen clearly in silhouette.  This may be related to
the fact that the silhouette disk appears somewhat asymmetric, having
a size that is $\sim$50\% larger toward the north at the same contour
level (Figure 2$b$).  The silhouette is also curved slightly, with the
concave opening toward the east, resulting from the contamination of
reflected light.

%% reflection
The shape of the silhouette and the bright reflection nebula on its
east-facing side indicate that the disk is flared (consistent with the
expected appearance in 2-D models; Whitney \& Hartmann 1992; Whitney
et al.\ 2003a, 2003b), reaching latitudes of $\pm$13$\arcdeg$ at its
northern and southern extremities, where the projected width is about
0.6\arcsec\ or 275 AU.  Our line of sight to the central star is
apparently very close to the flared surface of the disk, requiring
that we view the system at a latitude comparable to the flare angle;
thus, the disk axis is tilted by 10-15$\arcdeg$ out of the plane of
the sky, with the eastern portion facing toward us.  This viewing
geometry and the resulting asymmetric nebula are reminiscent of the
reflection nebula around VY CMa (Smith et al.\ 2001), except that the
disk of VY CMa is not seen in silhouette.  In the case of VY CMa, the
star is apparently viewed along a line of sight close to the surface
of the disk, and the position of the central star is wavelength
dependent (Smith et al.\ 2001; Kastner \& Weintraub 1998).  However,
in d216-0939, the central emission peak is the same in the F550M and
F658N filters (Figure 2$c$), suggesting either that the star is viewed
more directly, or that the dust grains in the protoplanetary disk are
large and do not strongly redden the visual-wavelength light (see the
discussion of the d114-426 disk by Shuping et al.\ 2003).

%% reflection - HH 30
The d216-0939 disk and its reflection nebula closely resemble the
well-studied HH~30 system in the L1551 dark cloud in Taurus (Burrows
et al.\ 1996; Stapelfeldt et al.\ 1999), although the physical
dimensions of this disk in Orion are several times larger than the
HH~30 disk.  This dark disk bisecting a bright reflection nebula is
reminiscent of several other disks as well (e.g., Padgett et al.\
1999; Stapelfeldt et al.\ 1998, 2003).  The brightness contrast
between the east and west halves in d216-0939 is stronger than in the
two halves of HH~30; this is consistent with our conjectured
inclination of the disk (with the axis tilted 10-15$\arcdeg$ from the
plane of the sky), since Burrows et al.\ (1996) derive a slightly
smaller tilt angle of $\sim$7$\fdg$5 (closer to edge-on) for HH~30.
As discussed below, the similarity of d216-0939 and HH~30 is even
further exemplified by their bipolar HH jets.

%%%% now talk about the jet -- mention HH30 again
\subsection{HH~667 -- A Bipolar Jet From d216-0939}

H$\alpha$ emission from HH~667 is presumably excited by shocks, but if
it is instead an irradiated jet, the discussion below remains valid.
This emission traces a bipolar outflow from the young star embedded in
the d216-0939 disk.  The ACS images reveal several knots near the
central star, as well as more distant features about 30\arcsec\ to the
east and west.

%% Microjet
{\it HH 667 microjet}: Figure 2 shows a system of emission knots along
an axis perpendicular to the disk plane, within 1$\arcsec$ from the
central peak.  One bright elongated knot is seen $\sim$0$\farcs$5 east
of the star, and two fainter emission knots are seen west of the star
along the outflow axis at 0$\farcs$4 and 0$\farcs$9 from the central
peak.  These are best seen in the contour map in Figure 2$b$ and the
intensity tracing in Figure 2$c$.  These knots are seen in H$\alpha$ +
[N~{\sc ii}] $\lambda$6583 line emission: they appear red/orange in
the color image in Figure 2$a$, and the tracings in Figure 3$c$
confirm that the H$\alpha$ emission from the knots has a clear excess
compared to scattered continuum light in the F550M filter (whereas the
central star in H$\alpha$ matches the continuum distribution).

This object is far from the center of M42 or M43, so if it is an
irradiated jet or dominated by shock excitation, it is likely that the
ionization fraction is low and that [N~{\sc ii}] $\lambda$6583 makes a
significant contribution to the emission in the F658N filter.
Therefore, we do not attempt to estimate the electron density from the
H$\alpha$ emission measure here.  High resolution spectra that could
separate these two emission lines and measure kinematics would be
useful for assessing the mass-loss rate in the HH~667 microjet.

%% HH 667 E
{\it HH 667 E}: The eastern portion of this flow is marked by a
partial bow shock located about 30\arcsec\ from the source.  This
feature is several arcseconds in extent and consists of two
sharp-edged filaments resembling bow shocks moving eastward (Fig.\
3$a$).  Very faint diffuse emission fills the interior of these bows
on their western sides.  The center of curvature of these partial bow
shocks is $\sim$11$\arcdeg$ off axis, suggesting that the flow may be
bent toward the south.

%% HH 667 W
{\it HH 667 W}: The western portion of the flow consists of several
faint and diffuse filaments of emission elongated along the outflow
axis, ranging from about 20\arcsec\ to 30\arcsec\ from the source (see
Fig.\ 3).  These features probably trace shocks within a faint jet.
The components of HH~667~W are less than 1\arcsec\ wide in the
direction perpendicular to the flow axis, but unlike the eastern part
of the flow, they are not significantly off-axis compared to the
bipolar microjet and disk of the central star.

The maximum separation between the eastern and western components of
the outflow from d216-0939 is 66\arcsec\ or about 0.15 pc in
projection.  Altogether, the d216-0939 disk and its more extended
HH~667 outflow closely resemble the well-studied HH~30 system in the
L1551 dark cloud in Taurus (Burrows et al. 1996; Stapelfeldt et
al. 1999).

%Finally, we note that d216-0939 may be a member of a multiple star
%system.  A second star located 4.3\arcsec\ (2000 AU) from the center
%of the disk drives a faint jet towards the north at an angle of about
%80$\arcdeg$ with respect to the flow from d215-939.

\section{d253-1536 AND THE HH~668 JET} 

Ground-based narrowband images of the H~{\sc ii} region M43 revealed a
compact globule with a central star embedded within the western rim of
this nebula.  The ACS images (Figure 4) show that the object is in
fact a proplyd ionization front or shock surrounding two stars
separated by 1$\farcs$1: the brighter star is visible on ground-based
images, and near it is a fainter star embedded in a silhouette disk.
The second object, d253-1536, drives a microjet visible on one side of
the disk, and the larger bipolar HH~668 jet that breaks-out on both
sides of the proplyd's bright rim in a direction perpendicular to the
disk.  A bow shock (HH~668~A) is seen almost 1$\arcmin$ away from the
central star to the south/southeast.  The various components of
d253-1536 and the HH~668 jet are summarized in Figure 4 and are
discussed below.

\subsection{The d253-1536 Disk and Bright Rim}

A 0$\farcs$6$\times$1$\farcs$4 disk seen in silhouette against
background nebular emission of M43 is located 1$\farcs$1 due east of
the brighter star.  Assuming that the d253-1536 disk is circular, the
axis of the disk is inclined by $\sim$25$\arcdeg$ with respect to the
plane of the sky.  However, the disk has pronounced asymmetry.  The
brightest pixel, presumably produced by a compact reflection nebula
located somewhat south of the star's true location, is displaced from
the disk center by about 0$\farcs$15 -- 0$\farcs$2 toward the east.
This distortion provides evidence that the bright star and the disk
form a true physical pair in which the gravitational perturbations of
the brighter and presumably more massive star generate the disk
asymmetry.  The d253-1536 disk is remarkable for being the secondary
member of a binary system in which the primary is a brighter star that
shows no indication of being surrounded by a visible disk.  The
secondary is decentered in its own circumstellar disk, but powers a
remarkably straight bipolar jet (see below), suggesting that the
orbital period is much longer than the flow timescale in the visible
jet.  Thus, the orbital timescale must be $>$100 yr.  For the apparent
separation of 1$\farcs$1 or a=$\simeq$250 AU, we find that the orbital
period would be $\sim$4,000 (M/M$_{\odot}$)$^{-0.5}$ yr, where M is
the total mass of the binary system.  These are not tight constraints
on the system's parameters, but at least they are consistent.  There
is no evidence for a larger circumbinary disk around the system.

Both members of this binary system are embedded within a proplyd
ionization front that faces directly toward NU Orionis, the early
B-type exciting star of the M~43 H~{\sc ii} region.  The bright rim's
radius is about 3$\arcsec$ (1400 AU), so the diameter of this object
is comparable to some of the giant proplyd candidates seen recently in
the Carina nebula (Smith et al.\ 2003).  The ionization front is
highly structured and consists of a series of scalloped rims, perhaps
signifying instabilities in a shock front rather than a smooth
ionization front.  An anonymous referee pointed out that this feature
bears a remarkable resemblance to the bright cusp seen around the dark
proplyd d114-426 (O'Dell \& Beckwith 1997).

\subsection{The HH 668 Jet}

A faint microjet (see \S 5.3) emerges from the center of the disk on
its south side, and its faint emission can be traced to the edge of
the proplyd's bright rim about 4$\arcsec$ to the south (Fig.\ 4$b$).
Then, as it breaks out of the proplyd ionization front, the jet
becomes brighter (HH~668~S) and can be traced for another 8$\arcsec$
toward the south along the same axis, which is perpendicular to the
silhouette disk.  Although the jet blends into the background nebular
light of M43 at this point, the flow continues farther to the south,
since a bright bow shock is located 55$\arcsec$ south on the jet axis
(HH~668~A in Figs.\ 4$a$, $c$, and $d$).  This south-facing bow has a
compact 0$\farcs$5 -- 1$\arcsec$-diameter apex located at
$\alpha$(2000) = 5$^h$ 35$^m$ 26.$^s$6, $\delta$(2000) = --5\arcdeg\
16\arcmin\ 25\arcsec, with an extended wing of H$\alpha$ emission that
trails back toward the source along its western rim for about
10$\arcsec$ (Figs.\ 4$a$ and $c$).  This feature is probably the
trailing edge of the bow shock.  HH~668~A has a width of only several
tenths of an arcsecond and has been missed on ground-based images.
The head of the bow shock is also bright in [O~{\sc iii}] (Fig.\
4$d$), indicating shock speeds of $\sim$100 km s$^{-1}$.

A counter-jet (HH~668~N) emerges from the northern side of the object
and can be traced out to at least 10\arcsec\ from the center of the
silhouette disk (Fig.\ 4$b$).  Thus, the bipolar HH~668 jet from the
disk in this binary star system has a projected spatial extent of at
least 65\arcsec\ (0.14 pc), similar to HH~667.  A line connecting
HH~668~N and S passes directly through the center of the d253-1536
disk, and comes within a few degrees of the HH~668~A bow shock.  Thus,
we do not detect major deflection of the jet by a large scale wind in
M43 or precession of the jet due to orbital motion in the binary
system, as mentioned above in \S 5.1.  However, proper motions do
reveal some deviations from a purely linear jet (see \S 5.4).  The N
and S components of the HH~668 jet (Fig.\ 4$b$) are brightest
immediately outside the proplyd's bright cusp, implying that at least
in this portion of the flow, HH~668 is an irradiated jet.

\subsection{The HH~668 Microjet From d253-1536}

Figure 5 shows details of the HH~668 microjet emerging from d253-1536.
Like most of the nearly two dozen microjets from proplyds in Orion
(Bally et al.\ 2000), it appears to be a one-sided jet where the
counter-jet is faint and lost in the bright background nebular light.
The jet is brightest on the southeast side of the disk, which may be
more directly illuminated by NU~Ori in the center of M43.

We obtained F658N images of d253-1536 during both Cycles 11 and 12
(2002 August and 2004 January, respectively), allowing us to document
temporal changes in the microjet structure.  Clear qualitative
differences are seen in Figures 5$a$ and 5$b$: in 2002 August (Fig.\
5$a$) the HH~668 microjet shows one bright emission knot located
roughly 0$\farcs$2 left of the star (note that the images in Fig.\ 5
are rotated so that the horizontal direction is along
P.A.=170$\arcdeg$), while 513 days later in 2004 January (Fig.\ 5$b$)
two distinct knots are seen.  If we assume that the second knot
farther from the star in 2004 January is the same condensation as the
single knot seen in 2002 August, then the centroid of this
condensation has moved 0$\farcs$195, indicating a transverse velocity
in the plane of the sky of roughly 300 km s$^{-1}$.  With the
d253-1536 disk axis tilted $\sim$25$\arcdeg$ from the plane of the sky
(see \S 5.1), the true speed of the HH~668 microjet is roughly 330 km
s$^{-1}$.  This is on the high end compared to typical microjet
speeds, and is significantly faster than downstream portions of the
HH~668 jet (see Table 3 and \S 5.4), implying that the flow
decelerates when it interacts with the ambient medium or downstream
jet material, as expected in conventional HH flows.

The observed H$\alpha$ surface brightness can be used to estimate the
electron density of the knots in the HH~668 microjet, with the
important caveats that this will be an underestimate if the jet is not
fully ionized and an overestimate if there is significant
contamination by [N~{\sc ii}] $\lambda$6583 in the filter.\footnote{We
made no attempt to correct for the uncertain contribution of [N~{\sc
ii}] $\lambda$6583, as the required information is not yet available.
To flux calibrate the images, we simply followed the standard
calibration method for {\it HST} imaging data -- i.e., using the {\sc
photflam} and {\sc photbw} values.}  The electron density is given by

\begin{displaymath}
n_e = \sqrt{ \frac{EM}{L_{\rm pc}} }
\end{displaymath}

\noindent where $EM$=4.89$\times$10$^{17}$ $I(H\alpha)$ is the
emission measure in cm$^{-6}$ pc (Spitzer 1978), and $I(H\alpha)$ is
the observed H$\alpha$ surface brightness in ergs s$^{-1}$ cm$^{-2}$
arcsec$^{-2}$.  $L_{\rm pc}$ is the typical emitting path length or
the diameter of the jet in pc.  From the tracings in Figure 5$c$ we
see that a typical value for $F_{\lambda}(H\alpha)$ in the knots is
$\sim$2$\times$10$^{-16}$ ergs s$^{-1}$ cm$^{-2}$ \AA$^{-1}$
arcsec$^{-2}$, or $I(H\alpha)\simeq$8$\times$10$^{-16}$ ergs s$^{-1}$
cm$^{-2}$ arcsec$^{-2}$ integrated over the 37.2 \AA\ effective width
of the F658N filter.  Adopting a jet width of $\sim$0$\farcs$1 or
$L$=46 AU as seen in images, we find $n_e\ga$4$\times$10$^3$
cm$^{-3}$.  (Again, this is probably a lower limit, since the
calculation does not include a correction for the ionization fraction
and because the jet width may be unresolved.)

Combining this electron density with the speed measured in proper
motions can yield an estimate of the mass-loss rate of the microjet
very close to the point of origin.  To do this, we must consider that
the observed emission knots may be either discrete blobs of gas, or
may be shocks in an otherwise steady flow.  In the first case, we must
average the electron density over the distance between the knots, so
likely values for the jet density are in the range 1--4$\times$10$^3$
cm$^{-3}$ (i.e., adopting an appropriate pseudo filling-factor).  The
average mass-loss rate is given by

\begin{displaymath}
\dot{M} = \pi \Big{(} \frac{L}{2} \Big{)}^2 m_H n_e v 
\end{displaymath}

\noindent yielding $\dot{M}\ga$3$\times$10$^{-10}$ ($n_e/10^3$)
M$_{\sun}$ yr$^{-1}$.  Since $n_e$=4$\times$10$^3$ cm$^{-3}$ is an
underestimate for the reasons stated above, the likely current mass
loss rate in the HH~668 microjet is of order 10$^{-9}$ M$_{\sun}$
yr$^{-1}$.

\subsection{Proper Motion of HH 668}

As noted above, we obtained {\it HST}/ACS F658N images of d253-1536
and HH~668 in both Cycle 11 and Cycle 12, allowing us to search for
proper motion in the components of the HH~668 jet using the
``difference-squared" method (e.g. Bally et al. 2002; Hartigan et al.\
2001) optimized for the measurement of nebular proper motions.  The
two sets of images were obtained 513 days (1.4 years) apart.  At a
distance of 460 pc assumed for this portion of the Orion nebula, a one
pixel (0$\farcs$05) shift in this time interval corresponds to a
velocity of about 75 km s$^{-1}$ .  On bright and compact features,
our image comparison method can determine motions to an accuracy of
about 0.1 to 0.3 pixels, or about 10 to 20 km s$^{-1}$.  The resulting
proper motions for components of the HH~668 jet are listed in
Table~3.\footnote{The bottom row of Table~3 lists the results of
measuring proper motion of a reference star, giving the method's
measurement uncertainty.} Figure 6 is similar to Figure 4, except that
the proper motions of various knots measured by this method are
over-plotted and subcomponents of the jet are labeled, corresponding to
the names in Table 3.  In Figure 6$a$, the arrows show the amount of
motion extrapolated to 50 years, while in the close-up view in Figure
6$b$, the arrows correspond to motion in 20 years.

The bipolar jet emerging from the silhouette disk exhibits motions
with velocities that decline from around 300 km s$^{-1}$\ for the
microjet within 1\arcsec\ of the central star (see above) to about 215
km s$^{-1}$\ at a distance 10\arcsec\ south.  A similar decline of
more than 30\% in the jet speed is seen toward the north.

The HH~668 flow can be traced about 50\arcsec\ south to a prominent
H$\alpha$ bow shock (HH~668A).  Faint wings of H$\alpha$ emission
sweep back along the western rim of this feature toward the source for
more than 15\arcsec .  A compact high-excitation region bright in
[O~{\sc iii}] $\lambda$5007 lies about an arcsecond upstream from the
tip of the shock.  The proper motion of the HH~668~A shock in
H$\alpha$ is about 140 km s$^{-1}$ toward the southeast; this speed is
sufficient to account for the excitation of the [O~{\sc iii}]
$\lambda$5007 emission line seen there.  Since HH~668~A is the only
part of the flow showing detectable [O~{\sc iii}] emission, this is
probably the terminal bow shock.  A fainter sub-arcsecond diameter
knot of high proper motion material is located between the source and
this bow shock about 40\arcsec\ south from the source (HH~668~B; see
Fig.\ 6$a$).  This knot shows proper motions of about 110 km s$^{-1}$
along the HH~668 flow axis, but has no [O~{\sc iii}] emission, so it
is an internal working surface of the jet.  A few other knots were
seen upon blinking images at the two epochs, but these did not yield
statistically significant results with our proper motion measurement
technique.  HH~668~A and B located relatively far from the source
support a general trend of a systematic decline in the proper motions
with increasing distance from the source.

The proper motion vectors show evidence for a systematic deflection of
the HH~668 bipolar outflow toward the east by about 10\arcdeg.  This
mild deflection is unusual in that it is toward the source of ionizing
radiation NU Ori. A possible explanation for this behavior is that the
motion of plasma within this portion of the M~43 H~{\sc ii} region is
dominated by the flow of material away from the western ionization
front located just beyond the right edge of Figure 4.

\section{CONCLUSIONS}

Our {\it HST}/ACS imaging survey of the outer parts of Orion and M43
yielded the discovery of 9 new disks seen in silhouette against faint
background light, and they show important new details for one
previously discovered silhouette disk.  Compared to other silhouette
disks, several members of this new sample are unusual in that they
show evidence for bipolar reflection nebulae or microjets.  This may
be a selection effect due to the fainter background levels far from
the Trapezium. Two of the new disks in our sample are found in or near
M43 and deserve special attention:

1.  d216-0939 is a nearly edge-on disk with a remarkable reflection
    nebula and a bipolar microjet that resembles the HH~30 disk/jet
    system in Taurus.  The disk is as large as the largest
    previously-known silhouette disk.  It also drives the 0.15 pc long
    bipolar jet HH~667.

2.  d253-1536 is also a very large silhouette disk, and is the only
    silhouette disk known so far to exhibit tidal distortions induced
    by a companion star in a binary system.  It is located inside the
    M43 H~{\sc ii} region, and is associated with a bright proplyd
    ionization front.  The disk drives both a microjet and a larger
    0.14 pc long bipolar jet called HH~668.  Images taken during
    Cycles 11 and 12 reveal detectable proper motions in the jet, with
    de-projected speeds as high as 330 km s$^{-1}$.  With an electron
    density of order 4$\times$10$^3$ cm$^{-3}$ near the source, this
    implies a mass-loss rate of roughly 10$^{-9}$ M$_{\sun}$
    yr$^{-1}$.

\acknowledgments  \scriptsize

We thank an anonymous referee for a timely and helpful review.  HH
catalog numbers are assigned by B.\ Reipurth in order to correspond
with the list of Herbig-Haro objects that he maintains (see {\url
http://ifa.hawaii.edu/reipurth/}).  Support was provided by NASA
through grants GO-9460, GO-9825, and HF-01166.01A from the Space
Telescope Science Institute, which is operated by the Association of
Universities for Research in Astronomy, Inc., under NASA contract
NAS5-26555.

% REFERENCES

% TABLES
\begin{deluxetable}{lccll}
%\tabletypesize{\normalsize}
\tabletypesize{\scriptsize}
\tablecaption{Cycle 11 ACS/WFC Observations}
\tablewidth{0pt}
\tablehead{
\colhead{Object} & \colhead{$\alpha_{2000}$} & \colhead{$\delta_{2000}$}
 & \colhead{Filter} 
 & \colhead{Exposure} \\ 
 \colhead{} & \colhead{} & \colhead{} & \colhead{} & \colhead{(sec)} 
}
\startdata
M43-NW    & 5 35 22.6 &  -5 16 09 & F658N & 2 $\times$ 500 \\
M43-NW    & 5 35 22.6 &  -5 16 09 & F660N & 2 $\times$ 500 \\
M43-NW    & 5 35 22.6 &  -5 16 09 & F550M & 30 \\
M43-NW    & 5 35 22.6 &  -5 16 09 & F502N & 4 $\times$ 500 \\
M42 North & 5 35 20.5 &  -5 11 17 & F658N & 2 $\times$ 500 \\
M42 North & 5 35 20.5 &  -5 11 17 & F660N & 2 $\times$ 500 \\
M42 North & 5 35 20.5 &  -5 11 17 & F550M & 30 \\
M42 North & 5 35 20.5 &  -5 11 17 & F502N & 4 $\times$ 500 \\
\enddata
%% Text for table notes should follow after the \enddata but before
%% the \end{deluxetable}. Make sure there is at least one \tablenotemark
%% in the table for each \tablenotetext.
\end{deluxetable}

\begin{deluxetable}{lcccl}
%\tabletypesize{\normalsize}
\tabletypesize{\scriptsize}
\tablecaption{New Silhouette Disks in Orion and M43}
\tablewidth{0pt}
\tablehead{
 \colhead{Object} &\colhead{$\alpha_{2000}$} &\colhead{$\delta_{2000}$}   
 &\colhead{Diameter}  &\colhead{Comments}
}
\startdata
d053-717  &5 35 05.41 &-5 27 17.2 &0$\farcs$9 &small disk, bright star \\
d110-3035 &5 35 10.99 &-5 30 35.2 &0$\farcs$9 &bipolar jet/refl.\ neb. \\
d124-132  &5 35 12.38 &-5 21 31.5 &0$\farcs$5 &proplyd, jet? \\
d132-042  &5 35 13.24 &-5 20 41.9 &0$\farcs$5 &proplyd, jet \\
d132-1832 &5 35 13.24 &-5 18 33.0 &1$\farcs$0 &variable star, swept disk \\
d141-1952 &5 35 14.05 &-5 19 52.1 &0$\farcs$7 &small disk, bright star \\
d216-0939 &5 35 21.57 &-5 09 38.9 &2$\farcs$6 &giant, edge-on, refl.\ neb., HH~667 \\
d253-1536 &5 35 25.30 &-5 15 35.5 &1$\farcs$5 &giant proplyd, binary, sil.\ disk, HH~668 \\
d280-1720 &5 35 28.04 &-5 17 20.2 &0$\farcs$8 &small disk, bright star \\
d347-1535 &5 35 34.67 &-5 15 34.8 &0$\farcs$7 &bipolar jet/refl.\ neb. \\
\enddata 
%% the \end{deluxetable}. Make sure there is at least one \tablenotemark
%% in the table for each \tablenotetext.
% \tablenotetext{a}{Sample footnote for table~\ref{tbl-1} 
% that was generated with the deluxetable environment}
\end{deluxetable}

% Table 3
\begin{deluxetable}{lcccccccccl}
\tabletypesize{\scriptsize}
\tablecaption{Proper Motions of the HH~668 Jet }
\tablewidth{0pt}
\tablehead{
\colhead{Name}             &
\colhead{$\alpha $(J2000)} &
\colhead{$\delta $(J2000)} &
\colhead{X}                &
\colhead{Y}                &
\colhead{D}                &
\colhead{V}                &
\colhead{PA (V)}           &
\colhead{PA (S)}           &
\colhead{age}         \\   
\colhead{}                 &
\colhead{h m s}            &
\colhead{\arcdeg\ \arcmin\ \arcsec\ }  &
\colhead{(\arcsec )}       &
\colhead{(\arcsec )}       &
\colhead{(\arcsec )}       &
\colhead{(km s$^{-1}$ )}          &
\colhead{(\arcdeg )}       &
\colhead{(\arcdeg )}       &
\colhead{(years)}      \\  
}
\startdata

N3         &5 35 25.01 &-5 15 24.4 &-4.3 &11.1  &11.9 & 84.5$\pm$21.8 & 354.1 &338.8 &  307   \\
N2         &5 35 25.08 &-5 15 26.0 &-3.3 &9.4   &10.0 &275.1$\pm$8.4  & 340.3 &340.8 &   79   \\
N1         &5 35 25.14 &-5 15 28.8 &-2.4 &6.7   &7.1  &275.1$\pm$8.4  & 340.3 &340.5 &   55   \\
microjet   &5 35 25.31 &-5 15 35.9 &0.09 &-0.49 &0.5  &303.5$\pm$8.0  & 170.0 &170.0 &  3.6   \\
S1         &5 35 25.39 &-5 15 39.9 &1.4  &-4.4  &4.6  &245.6$\pm$6.0  & 160.2 &162.9 &   40   \\
S2         &5 35 25.43 &-5 15 41.5 &2.0  &-6.0  &6.3  &108.8$\pm$16.0 & 145.3 &161.6 &  126   \\
S3         &5 35 25.48 &-5 15 42.7 &2.7  &-7.2  &7.7  &178.2$\pm$16.0 & 159.7 &159.9 &   94   \\
B          &5 35 26.08 &-5 16 10.4 &11.7 &-34.9 &36.8 &111.8$\pm$14.0 & 160.4 &161.4 &  717   \\
A          &5 35 26.61 &-5 16 25.2 &19.6 &-49.7 &53.4 &143.3$\pm$8.4  & 154.3 &158.5 &  812   \\
ref.\ star &5 35 27.06 &-5 15 44.6 &26.2 &-9.2  &27.8 &  5.9$\pm$6.0  &\nodata&109.2 &\nodata \\ 

\tablecomments{ \scriptsize
Column 1: Name of each knot (see Figure 6).
Column 2 \& 3:  The J2000 coordinates of each box center.:
Column 4: The projected distance from d253-1536 along the R.A.\ direction in arc seconds.
Column 5: The projected distance from d253-1536 along the DEC direction in arc seconds.
Column 6: The projected distance from d253-1536 along the jet axis in arc seconds.
Column 7: The proper motion in km s$^{-1}$ .
Column 8: The position angle of the measured velocity vector.
Column 9: The position angle of a line drawn from d253-1536 to
  the center of the box.
Column 10: The dynamical age of the feature (D/V) assuming a distance
  of 460 pc (the microjet age corresponds to the Cycle 12 observation).
}
\enddata
\label{PMH}
\normalsize
\end{deluxetable}

% FIGURE 1 ABCDEF
\begin{figure}
\epsscale{0.9}
\plotone{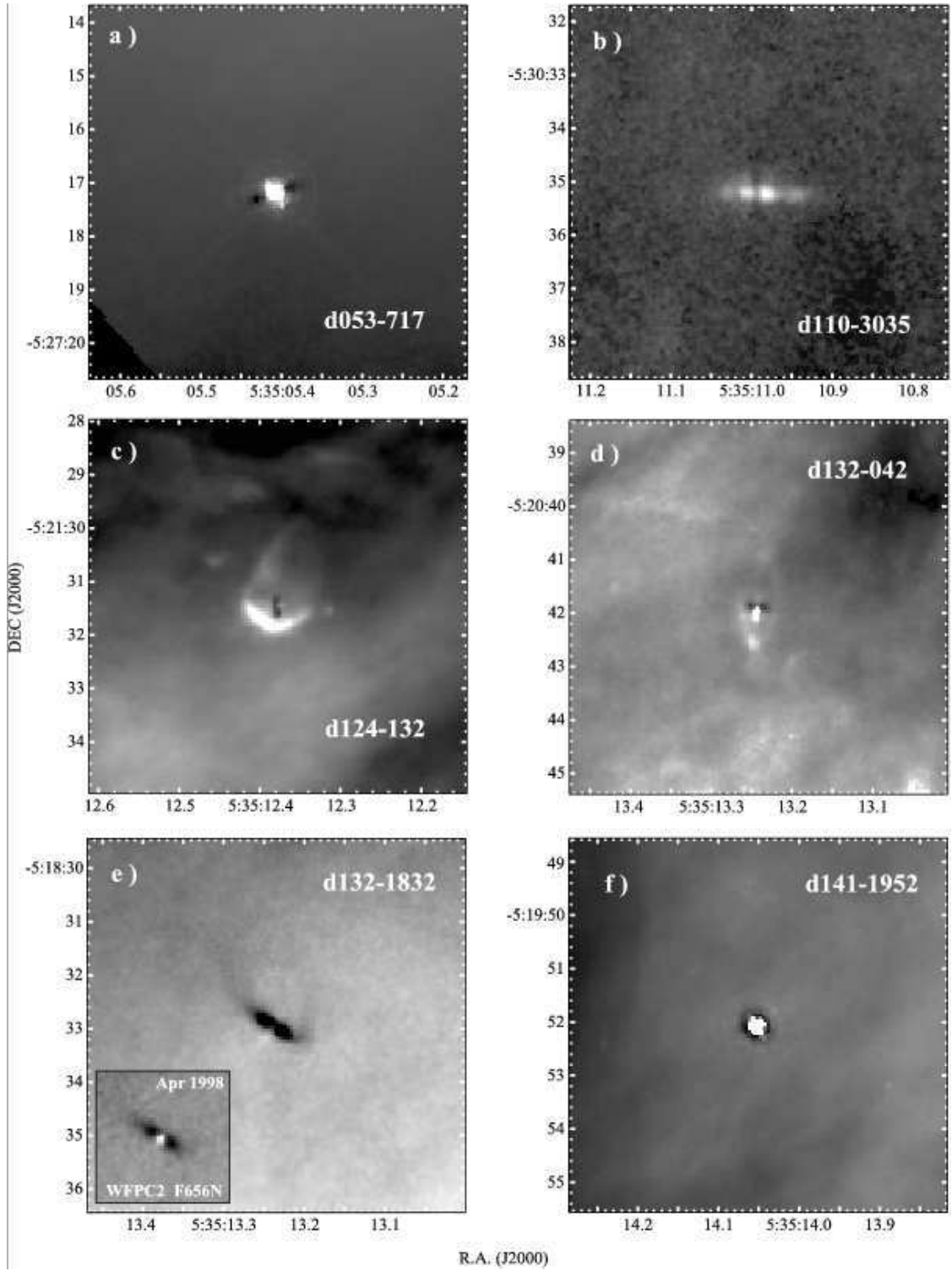}
\caption{ACS/WFC F658N (H$\alpha$) images of selected silhouette disks
in the outskirts of the Orion Nebula and M43.  All are
newly-discovered disks, except d132-1832 in panel (e), shown here
because of new structures seen in the outer parts of the disk and
because of the variable central star (inset; see text).}
\end{figure}

% FIGURE 1 GHIJ
\begin{figure}
\figurenum{1}
\epsscale{0.9}
\plotone{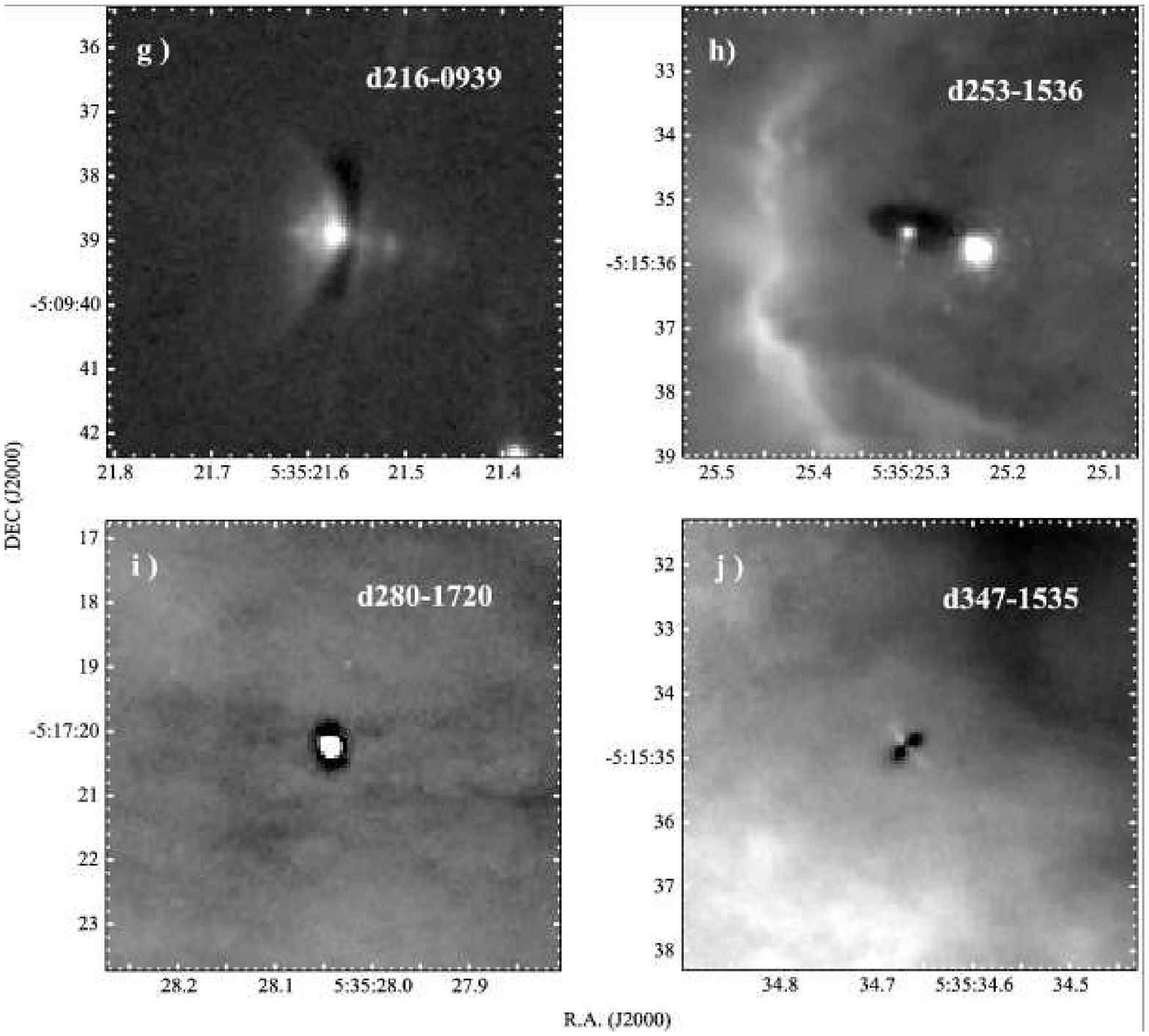}
\caption{continued.}
\end{figure}

% FIGURE 2
\begin{figure}
%\figurenum{1}
\epsscale{0.4}
\plotone{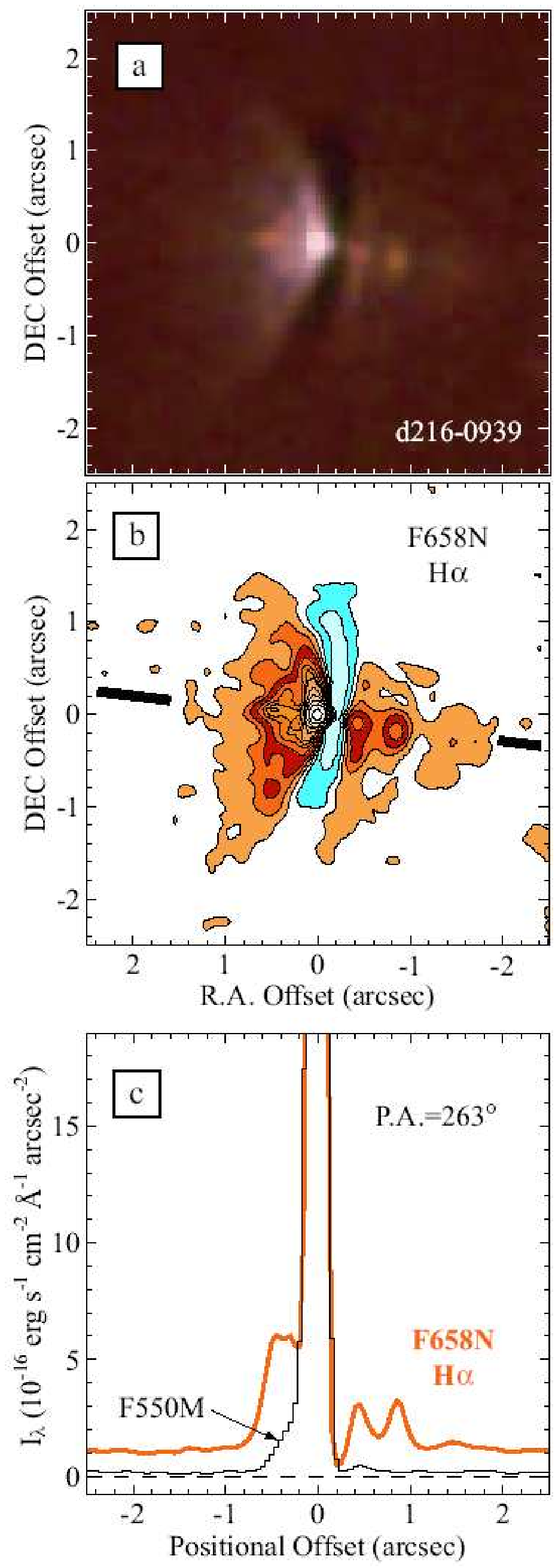}
\caption{(a) 3-color image of the reflection nebula d216-0939, with
F550M in blue, F660N in green, and F658N in red.  (b) Contour plot of
the deconvolved F658N H$\alpha$ image of d216-0939.  The two lowest
contours are blue and signify extinction of the background nebular
light, and are drawn at 0.5 and 0.7 $\times$10$^{-16}$ erg s$^{-1}$
cm$^{-2}$ \AA$^{-1}$ arcsec$^{-2}$.  The remaining contours are orange
and are emission above the background.  The lowest contour is roughly
3$\sigma$ above the background, and is drawn at 0.95
$\times$10$^{-16}$ erg s$^{-1}$ cm$^{-2}$ \AA$^{-1}$ arcsec$^{-2}$.
The remaining contours are drawn at 1.2, 1.6, 2.5, 3.7, 4.8, 6, 10,
20, 40, 80, and 200 $\times$10$^{-16}$ erg s$^{-1}$ cm$^{-2}$
\AA$^{-1}$ arcsec$^{-2}$. (c) Intensity tracings along
P.A.=263$\arcdeg$ through the central peak (marked by the black lines
in the middle panel) for the F658N (thick orange line) and F550M
(black histogram) filters.}
\end{figure}

% FIGURE 3
\begin{figure}
%\figurenum{1}
\epsscale{0.9}
\plotone{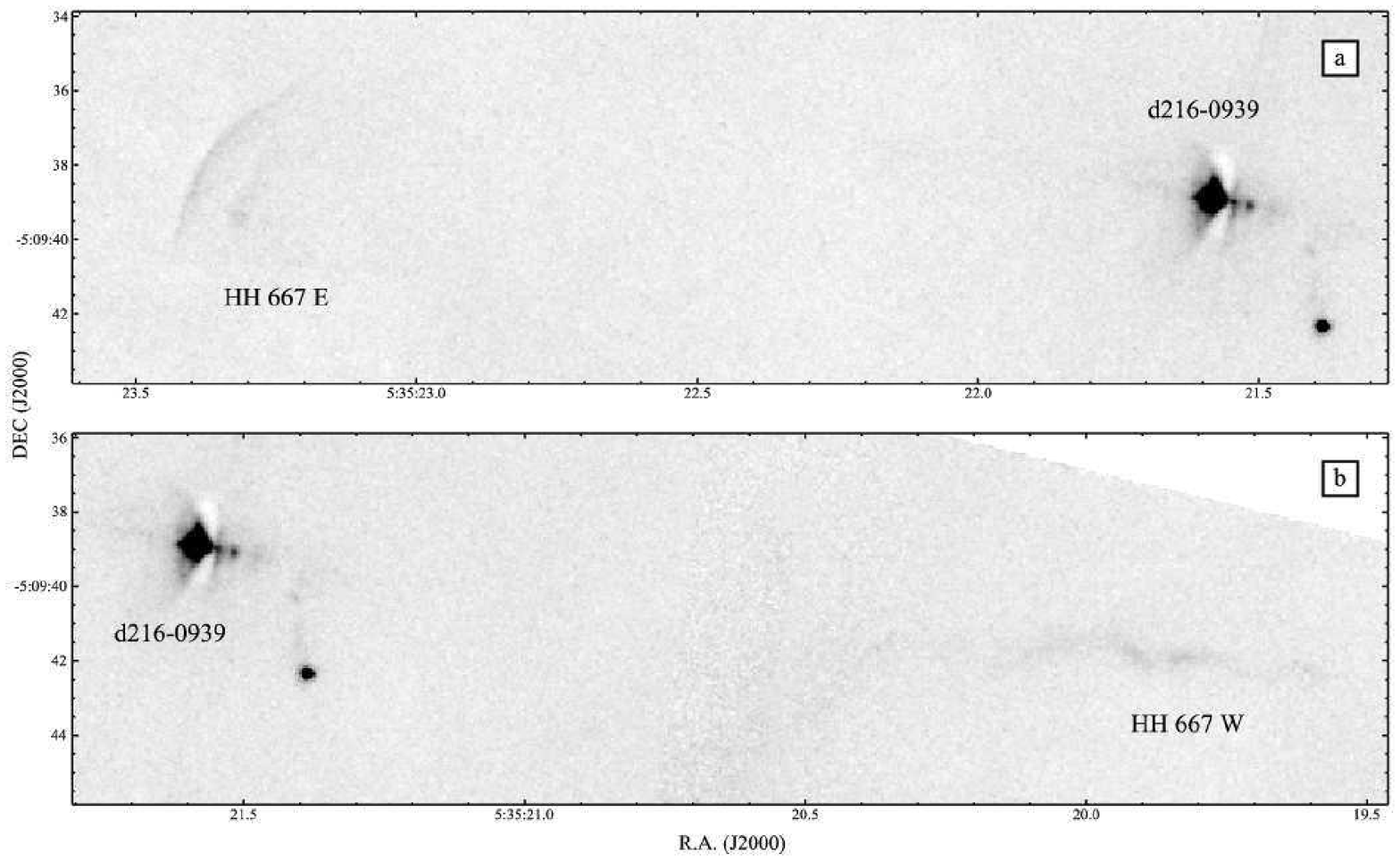}
\caption{HH 667, candidate bow shocks in the jet from d216-0939.  These
images are displayed in negative grayscale.  (a) the eastern portion
of the jet, with a curved bow shock (HH 667 E).  (b) the western
portion of the jet, with emission filaments along the jet axis (HH 667
W).}
\end{figure}

% FIGURE 4
\begin{figure}
%\figurenum{1}
\epsscale{0.9}
\plotone{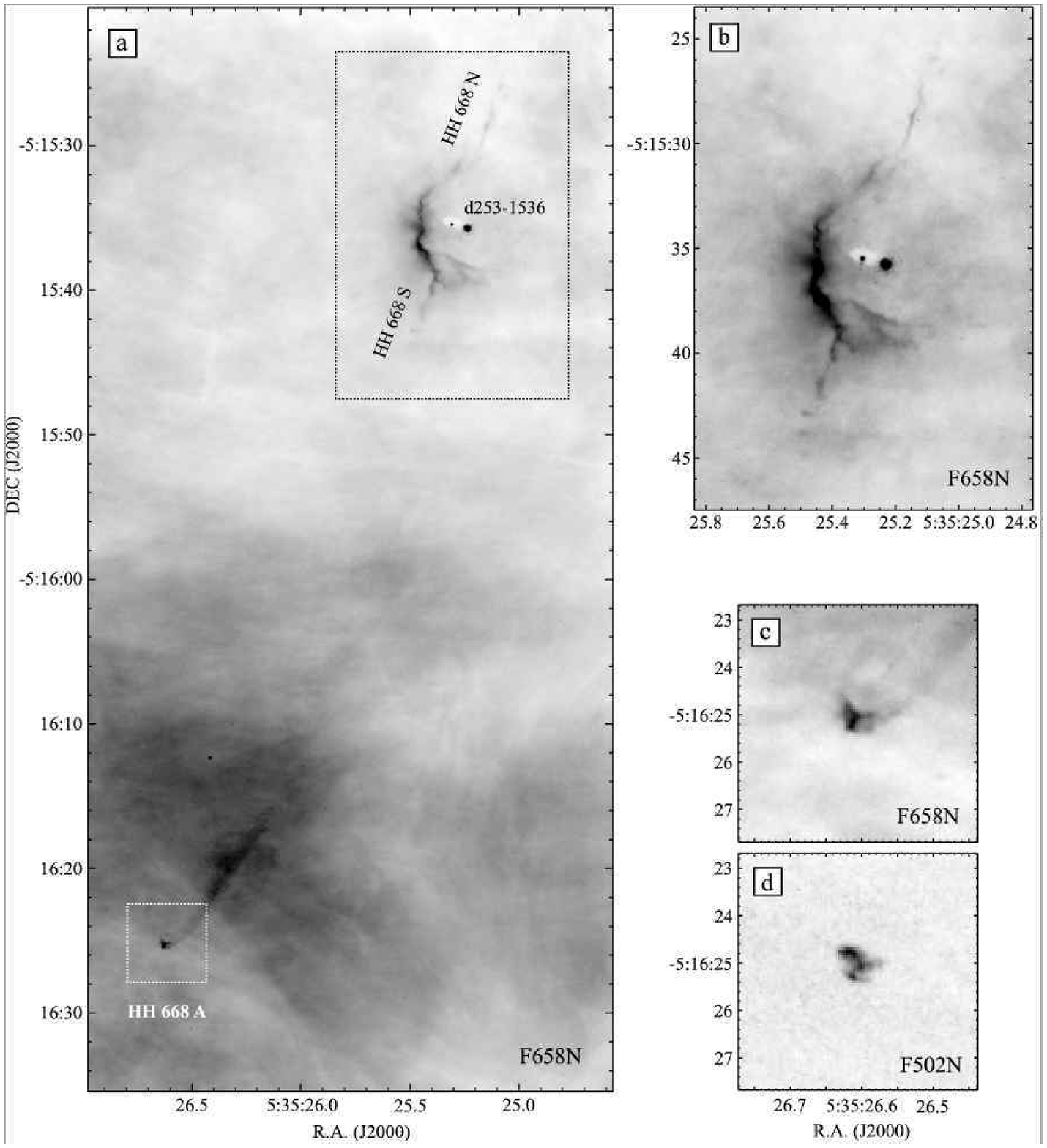}
\caption{Images of the binary proplyd d253-1536 and the HH~668 jet
with negative grayscale display.  (a) Large scale image showing the
alignment of the bipolar jet from d253-1536, which is perpendicular to
the disk, and the bow shock HH~668~A.  (b) detail of the d253-1536
proplyd and the HH~668 jet.  (c) and (d) are details of the HH~668~A
bow shock in H$\alpha$ (F658N) and [O~{\sc iii}] $\lambda$5007 (F502N)
emission, corresponding to the small white box in Panel $a$.}
\end{figure}

% FIGURE 5
\begin{figure}
%\figurenum{1}
\epsscale{0.45}
\plotone{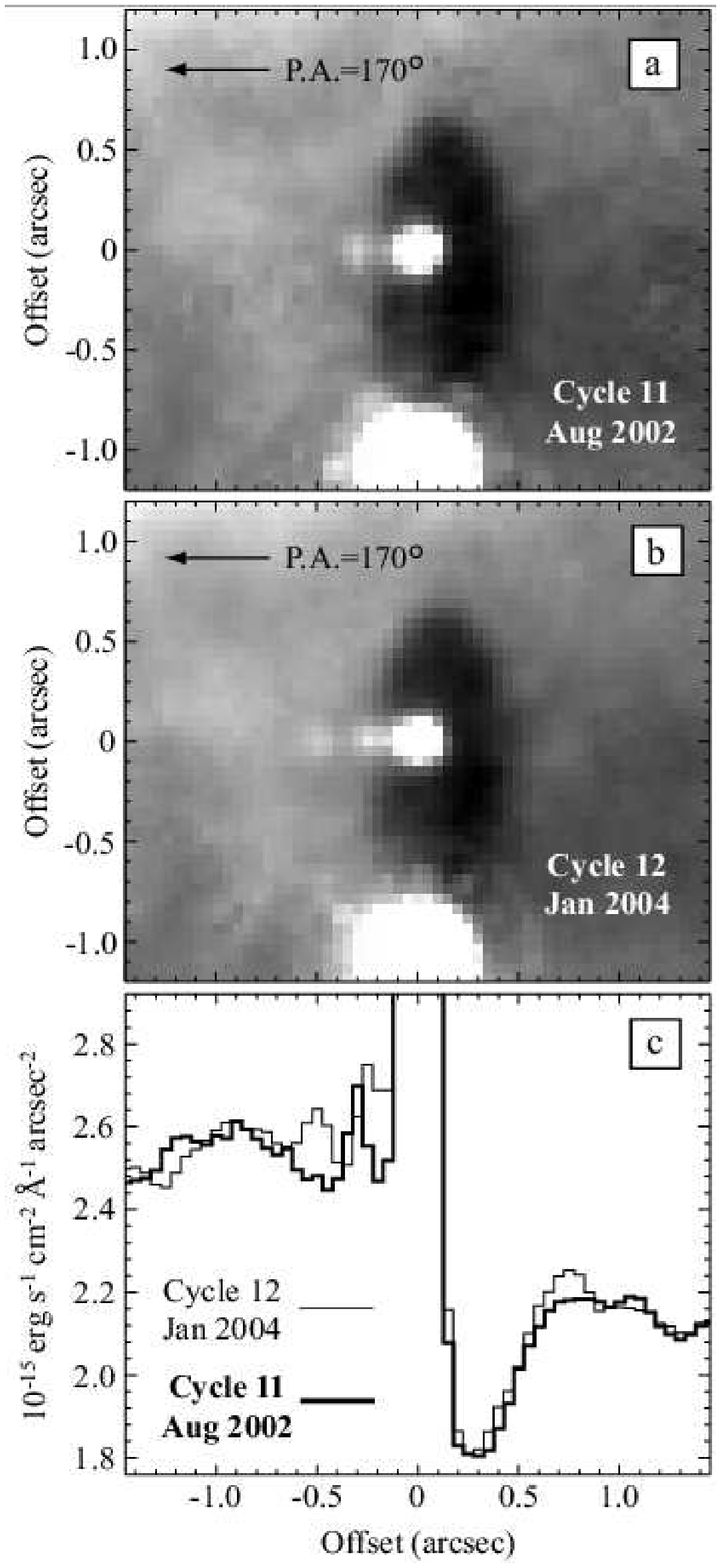}
\caption{The HH~668 microjet of d253-1536 at two epochs.  (a) Cycle 11
F658N image obtained in August 2002, rotated so that P.A.=170$\arcdeg$
is to the left.  (b) same as Panel $a$ but for Cycle 12 obtained in
January 2004.  (c) Intensity tracings of H$\alpha$ (F658N) emission
for both Cycles 11 and 12, passing through the star at
P.A.=170$\arcdeg$=350$\arcdeg$, showing the motion of the knots to the
left of the star in Panels $a$ and $b$.  A new knot has appeared in
Cycle 12, while the bright knot seen in Cycle 11 has moved.  Cycle 11
is the thicker line, and Cycle 12 is the thinner line.}
\end{figure}

% Fig 6
\begin{figure}
%\figurenum{1}
\epsscale{0.98}
\plotone{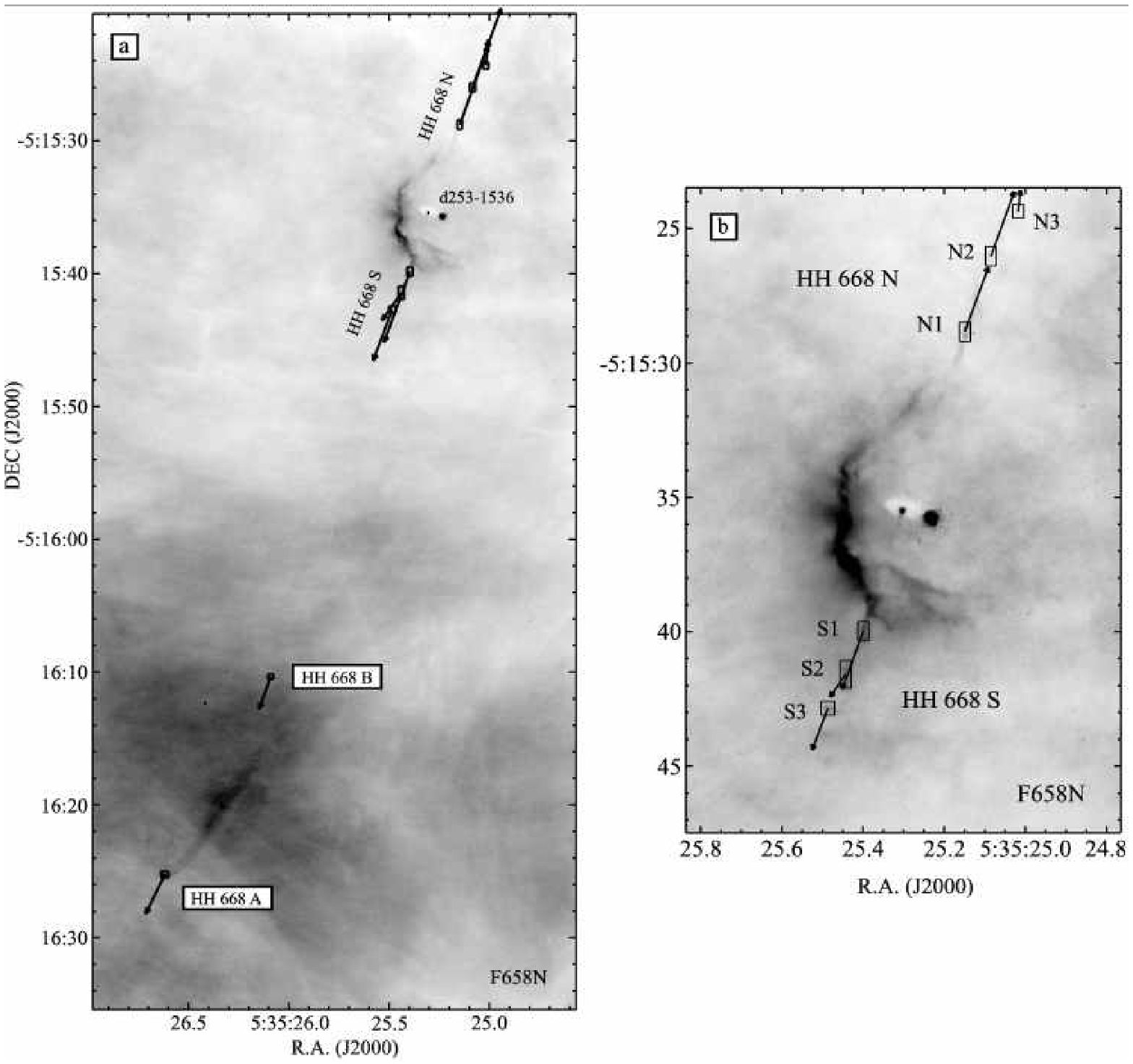}
\caption{Same as for Figure 4, but showing proper motion vectors for
various features as described in the text and listed in Table 3.
Panel (a) shows an expanded view of the HH~668 jet, with the vectors
corresponding to extrapolated motion in 50 yr, while panel (b) shows a
zoomed-in view of HH~668~N and S, where the vectors correspond to
extrapolated motion in 20 yr.}
\end{figure}


\begin{references}

\reference{} Bally, J., Heathcote, S., Reipurth, B., Morse, J.A.,
Hartigan, P., \& Schwarz, R.\ 2002, AJ, 123, 2627

\reference{} Bally, J., Licht, D., Smith, N., \& Walawender, J.\ 2004,
AJ, in press

\reference{} Bally, J., O'Dell, C. R., \& McCaughrean, M.\ 2000, AJ,
119, 2919

\reference{} Bally, J., \& Reipurth, B.\ 2001, ApJ, 546, 299

\reference{} Bally, J., Sutherland, R. S., Devine, D., \& Johnstone,
D.\ 1998, AJ, 116, 293

\reference{} Burrows, C.J., et al.\ 1996, \apj, 473, 437 

\reference{} Chen, H., et al.\ 1998, ApJ, 492, L173

\reference{} Hartigan, P., Morse, J.A., Reipurth, B., Heathcote, S.,
\& Bally, J.\ 2001, ApJ, 559, L157

\reference{} Hodapp, K.W., Walker, C.H., Reipurth, B., Wood, K.,
Bally, J., Whitney, B.A., \& Connelley, M.\ 2004, ApJ, 601, L79

\reference{} Kastner, J.H., \& Weintraub, D.A.\ 1998, AJ, 115, 1592

\reference{} McCaughrean, M.J., \& O'Dell, C.R.\ 1996, AJ, 111, 1977

\reference{} McCaughrean, M.J., et al.\ 1998, ApJ, 492, L157

\reference{} McCullough, P.R., et al.\ 1995, ApJ, 438, 394

\reference{} O'Dell, C.R., \& Beckwith, S.\ 1997 Science, 276, 1355

\reference{} O'Dell, C.R., \& Wong, S.-K.\ 1996, AJ, 111, 846

\reference{} O'Dell, C. R., Wen, Z., \& Hu, X.\ 1993, ApJ, 410, 696 

\reference{} O'Dell, C.R.\ 2001, \aj, 122, 2662 

\reference{} Padgett, D.L., Brandner, W., Stapelfeldt, K.R., Strom,
S.E., Terebey, S., \& Koerner, D.\ 1999, AJ, 117, 1490

\reference{} Reipurth, B., \& Bally, J.\ 2001, ARAA, 39, 403

\reference{} Shuping, R.Y., Bally, J., Morris, M., \& Throop, H.\
2003, ApJ, 587, L109

\reference{} Smith, N., Bally, J., \& Morse, J.A.\ 2003, ApJ, 587,
L105

\reference{} Smith, N., Barb\'{a}, R.H., \& Walborn, N.R.\ 2004,
MNRAS, 351, 1457

\reference{} Smith, N., Humphreys, R.M., Davidson, K., Gehrz, R.D.,
Schuster, M.T., \& Krautter, J.\ 2001, AJ, 121, 1111

\reference{} Spitzer, L.\ 1978, Physical Processes in the Interstellar
Medium (New York: Wiley)

\reference{} Stapelfeldt, K.R., et al.\ 1999, \apjl, 516, L95 

\reference{} Stapelfeldt, K.R., Krist, J.E., Menard, F., Bouvier, J.,
Padgett, D.L., \& Burrows, C.J.\ 1998, ApJ, 502, L65

\reference{} Stapelfeldt, K.R., Menard, F., Watson, A.M., Krist, J.E.,
Dougados, C., Padgett, D.L., \& Brandner, W.\ 2003, ApJ, 589, 410

\reference{} St\"{o}rzer, H., \& Hollenbach, D.\ 1998, ApJ, 502, L71

\reference{} Whitney, B.A., \& Hartmann, L.\ 1992, ApJ, 395, 529

\reference{} Whitney, B.A., Wood, K., Bjorkman, J.E., \& Cohen, M.\
2003a, ApJ, 598, 1079

\reference{} Whitney, B.A., Wood, K., Bjorkman, J.E., \& Wolff, M.J.\
2003b, ApJ, 591, 1049


\end{references}
\end{document}